\documentclass{jpsj2}
\title{Multicritical Points of Potts Spin Glasses on the Triangular Lattice}

\author{Masayuki Ohzeki}

\inst{Department of Physics, Tokyo Institute of Technology,
Oh-okayama, Meguro-ku, Tokyo 152-8551, Japan}

\abst{We predict the locations of several multicritical points of the Potts spin glass model on the triangular lattice. In particular, continuous multicritical lines, which consist of multicritical points, are obtained for two types of two-state Potts (i.e., Ising) spin glasses with two- and three-body interactions on the triangular lattice.
These results provide us with numerous examples to further verify the validity of the conjecture, which has succeeded in deriving highly precise locations of multicritical points for several spin glass models.
The technique, called the direct triangular duality, a variant of the ordinary duality transformation, directly relates the triangular lattice with its dual triangular lattice in conjunction with the replica method.}

\kword{spin glass, multicritical point, Potts model, exact solution, triangular lattice, duality transformation}
\begin{document}
\maketitle

\section{Introduction}
Numerous investigations and analyses have been carried out for many years to study the properties of finite-dimensional spin glasses.
While there have been many studies based on mean-field analyses, approximate approaches, and numerical simulations, there are much fewer exact or rigorous approaches \cite{EA,SK,Young}.
Among these studies for finite-dimensional spin glasses, a reliable approach has been developed to predict the exact location of the multicritical point. we call this approach the conjecture.
Since its proposal, many predictions for the locations of the multicritical points have been given and verified by numerical studies, including the $\pm J$ Ising model on the square\cite{NN,MNN}, triangular, and hexagonal lattices\cite{TSN,NO}, and various mutually dual pairs of hierarchical lattices\cite{HB}.

While investigation of the multicritical point is an issue in the phase diagram of spin glass models, it has also been pointed out that the location of the multicritical point of the $\pm J$ Ising model is closely related to an accuracy threshold of reliability of the toric code, one of the error-correcting codes used in quantum information\cite{DKLP}. 
It is therefore an important problem to determine the location of the multicritical point, not only for its intrinsic interest, but also for its influence in other branches of physics.
Most of the predicted results are in good agreement with numerical estimations, within their error bars.
However, the validity of the conjecture is still uncertain, and small deviations have been observed in a few samples\cite{HB}.
We must therefore further verify the validity of the conjecture for many examples and clarify the required conditions for obtaining the exact location of the multicritical point.
This is the goal of the present work.

Here, we introduce the Potts spin glass model with two- and three-body interactions.
The pure, non-random Potts model with two- and three-body interactions satisfies self-duality under the direct triangular duality\cite{Wu}.
Because this model has two independent coupling parameters, the phase boundary is described as a line in the two-dimensional phase diagram.
Similarly to this pure Potts model, the Potts spin glass model may have a line of multicritical points.
Based on these predictions, we investigate numerous examples to check the validity of the conjecture.
It may also be possible to investigate in detail the required conditions for obtaining the exact location of the multicritical point and to find the reason why some discrepancies have been observed.

This paper is organized as follows. 
We first review the properties of the pure Potts model with two- and three-body interactions on the triangular lattice in \S 2.
We introduce, in \S 3, the two types of Potts spin glass models with two- and three-body interactions on the triangular lattice.
In \S 4, we discuss the duality for these Potts spin glass models and the possibility of predicting their multicritical points.
We show that several cases of the Potts spin glass models satisfy the known conditions for obtaining the exact location of the multicritical point: the $2$-state Potts (i.e., Ising) spin glass models, and the $q$-state Potts spin glass models only with three-body interactions. 
We show the results of the presented analyses in \S 5. The results predict multicritical lines, consisting of multicritical points, for the two types of Ising spin glass models and multicritical points for the $q$-state Potts spin glass model only with three-body interactions.
Finally, we discuss the present work.
\section{Potts model with two- and three-body interactions}
The Potts model with two- and three-body interactions on the triangular lattice is a generalization of the ordinary Potts model with only two-body interactions on the triangular lattice.
It was first proposed by Kim and Joseph through the duality and star-triangle transformations for the ordinary Potts model on the triangular lattice, which has only two-body interactions on each bond\cite{KJ}.
Subsequently, the direct triangular duality was developed by simultaneous use of the duality and star-triangle transformations\cite{Wu}.
This direct triangular duality was applied to the Potts model with two- and three-body interactions as a general case of the analysis by Kim and Joseph.
It will be helpful to first review the pure $q$-state Potts model with two- and three-body interactions on the triangular lattice before we consider the random systems.

\begin{figure}[tb]
\begin{center}
\scalebox{1.4}{\includegraphics*[85mm,130mm][130mm,170mm]{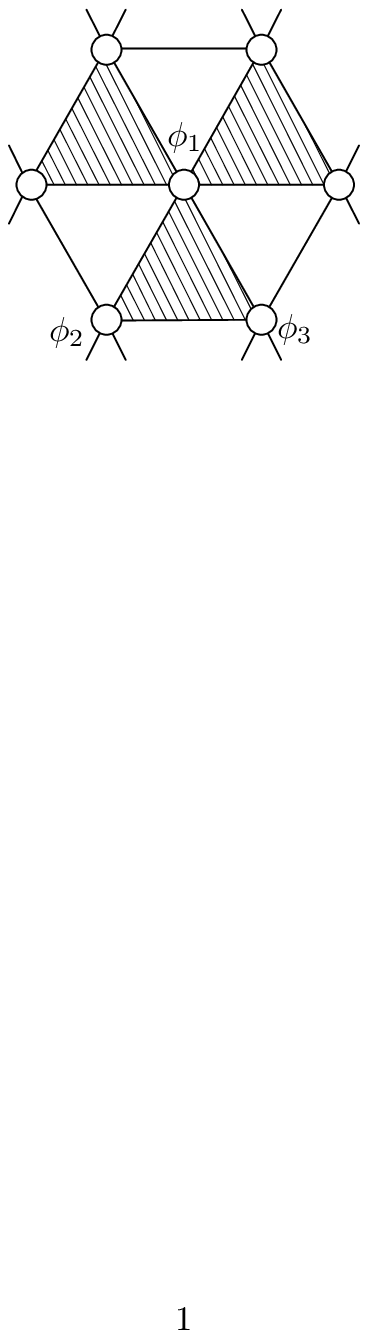}}
\end{center}
\caption{{\small The Potts model with two- and three-body interactions on a triangular lattice. Each of the spin variables interacts with nearest-neighboring sites
(two-body interactions). We also consider three-body interactions among spins
on all up-pointing triangles, shown as shaded. No three-body interactions are considered on the down-pointing triangles.}
}
\label{tri}
\end{figure}

We denote the spin variables as $\phi_i$ and the differences as $\phi_{ij} \equiv \phi_i - \phi_j$, which take integer values from 0 to $q-1$ (mod $q$). Then, the partition function of the Potts model with two- and three-body interactions is written as
\begin{equation}
Z = \sum_{\{ \phi_{i} \} }\prod_{\bigtriangleup} A_{\bigtriangleup} [\phi_{12},\phi_{23},\phi_{31}].
\end{equation}
Here, the product over $\bigtriangleup$ runs over up-pointing shaded triangles shown in Fig. \ref{tri}, and $A[\cdots]$ is the face Boltzmann factor on each up-pointing triangular face:
\begin{equation}
A^{K_2,K_3}_{\bigtriangleup}[\phi_{12},\phi_{23},\phi_{31}] \equiv
\exp 
\left\{
K_2 \sum_{i\neq j} \delta(\phi_{ij})
+
K_3 \prod_{i\neq j} \delta(\phi_{ij})
\right\},\label{A_ori}
\end{equation}
where $i\neq j$ runs over the bonds surrounding each up-pointing triangle (in particular, (12), (23), and (31) in Fig. \ref{tri}). $K_2$ and $K_3$ are coupling constants of the two- and three-body interactions. Thus, we describe the thermodynamic state as a point on the $K_2$-$K_3$ plane.
In the usual definition of the partition function, $K_2$ and $K_3$ are proportional to the inverse temperature $\beta = 1/k_B T$. From this viewpoint, a line is described on the $K_2$-$K_3$ plane by a change in temperature. This line is called the thermodynamic line\cite{Wu3}. If the thermodynamic line goes across the phase boundary, the Potts model undergoes a phase transition.

Next, we consider how to determine where the phase transitions occur.
The duality transformation is a useful tool to give the exact location of the transition point.
However, an ordinary duality, which is carried out by the Fourier transformation\cite{WuWang}, changes the structure of the lattice. For example, in our problem, the triangular lattice is transformed into a hexagonal lattice.
Therefore, unlike the simple case of the pure Ising model on the square lattice, it is necessary to apply another transformation, the star-triangle transformation, to relate the hexagonal lattice with another triangular lattice.
In this paper, we call the simultaneous use of these techniques the direct triangular duality.
It is carried out by application of the following transformation:
\begin{eqnarray}
   A^*[k_{12}, k_{23}, k_{31}]&=&\frac{1}{q}\sum_{\{\phi_{ij}\}}
 \delta (\phi_{12}+\phi_{23}+\phi_{31})A[\phi_{12},\phi_{23},\phi_{31}]
  \nonumber\\
 &&\times\exp\left\{ \frac{2\pi i}{q}
(k_1\phi_{23}+k_2\phi_{31}+k_3\phi_{12})\right\},
\label{A_dual}
\end{eqnarray}
where $A$ and $A^*$ are the original and dual face Boltzmann factors, respectively.
The summation with the subscript $\{\phi_{ij}\}$ runs from $\{\phi_{12},\phi_{23},\phi_{31}\}=\{0,0,0\}$ to $\{q-1,q-1,q-1\}$.
The quantities $k_i$ are spin variables on the dual triangular lattice, and $k_{ij}$ represent their differences $k_i - k_j$.
In eq. (\ref{A_dual}), the exponential terms are produced by the ordinary duality, and Kronecker's delta arises from the application of the star-triangle transformation.

The face Boltzmann factor (\ref{A_ori}) can be expanded by the identity $\exp(K\delta(\cdots)) = 1 + v\delta(\cdots)$, where $v\equiv \exp(K)-1$, as follows:
\begin{equation}
A^{K_2,K_3}_{\bigtriangleup}[\phi_{12},\phi_{23},\phi_{31}] = 1 + v_2 \sum_{i\neq j}\delta(\phi_{ij}) + y \prod_{i\neq j}\delta(\phi_{ij}).\label{23P1}
\end{equation}
Here, $y \equiv 3v_2^2 + v^3_2 + v_3(1+v_2)^3 = \exp(K_3+3K_2)-3\exp(K_2)+2$, $v_2 \equiv exp(K_2)-1$, and $v_3 \equiv exp(K_3)-1$.
By the direct triangular duality (\ref{A_dual}), the face Boltzmann factor is transformed to
\begin{equation}
A^{K^*_2,K^*_3}_{\bigtriangleup}[k_{12},k_{23},k_{31}] = \frac{y}{q}
\left\{1 + \frac{qv_2}{y} \sum_{i\neq j}\delta(k_{ij}) + \frac{q^2}{y} \prod_{i\neq j}\delta(k_{ij})\right\}.\label{23P2}
\end{equation}
This equation states that the Potts model with two- and three-body interactions is a self-dual model which satisfies
\begin{equation}
v_2^* = \frac{qv_2}{y}, \quad
y^* = \frac{q^2}{y}.\label{dr}
\end{equation}
The fixed-point conditions are therefore $v_2 = v_2^*$ and $y=y^*$.
Both of these fixed-point conditions give a single equation, $y=q$.
This condition $y=q$ gives a line consisting of the fixed points.
This fixed line is expected to determine a part of the phase boundary of the Potts model with two- and three-body interactions.
Wu and Zia\cite{Wu3} argued that this fixed line indeed is a phase boundary when the couplings satisfy the criteria $J_3+3J_2>0$ and $J_3+2J_2>0$.
These criteria mean that coupling constants $K_2$ and $K_3$ make effective ferromagnetic interactions; i.e., all the spins should be parallel in the ground state.

When $v_2 = 0$, the duality relations (\ref{dr}) are reduced to
\begin{equation}
v_3^* = \frac{q^2}{v_3}.
\end{equation}
In other words, the duality relations (\ref{dr}) map the Potts model with two- and three-body interactions $(v_2 \neq 0, v_3)$ into another Potts model, that is, the dual model, with two- and three-body interactions $(v^*_2 \neq 0, v^*_3)$.
If we apply the direct-triangular duality to the Potts model with only three-body interactions $(v_2=0, v_3)$, it is transformed into the dual Potts model with only three-body interactions $(v_2=0, v^*_3)$.
It is obvious that the fixed-point condition for the Potts model only with three-body interactions is $v_3=q$.

\section{Potts spin glass model}
In the previous section, we introduced the Potts model with two- and three-body interactions.
Here, we add quenched randomness to the two-body and three-body interactions in this Potts model.

The Hamiltonian of the Potts spin glass model on the triangular lattice is defined as
\begin{equation}
H = \sum_{\bigtriangleup}H_{\bigtriangleup}[\phi_{12},\phi_{23},\phi_{31}] = \sum_{\bigtriangleup}\left\{
-J_2 \sum_{i\neq j}\delta_(\phi_{ij}+l_{ij}) - J_3\prod_{i\neq j}\delta(\phi_{ij}+m_{ij})
\right\}\label{Ham},
\end{equation}
and its face Boltzmann factor is
\begin{equation}
A^{K_2,K_3}_{\bigtriangleup}[\{\phi_{ij}+l_{ij}\};\{\phi_{ij}+m_{ij}\}] = \exp
\left\{
K_2 \sum_{i\neq j}\delta_(\phi_{ij}+l_{ij}) + K_3\prod_{i\neq j}\delta(\phi_{ij}+m_{ij})
\right\}
,
\end{equation}
where $\phi_{ij}$ are the same symbols as in the previous section.
The quantities $l_{ij}$ are the random variables for the two-body interactions, and $m_{ij}$ are those for the three-body interactions.
These random variables follow the distribution functions
\begin{eqnarray}\nonumber
P(l_{ij}) &=&
\begin{cases}
 1-(q-1)p_2 & l_{ij} = 0\\ 
 p_2 & \textrm{otherwise}
\end{cases}\\
&=& p_2 \exp\{K_{p2}\delta(l_{ij})\} \label{Pl}\\ \nonumber
P(\{m_{ij}\}) &=&
\begin{cases}
 1-(q^3-1)p_3 & \{m_{ij}\}  = \{0,0,0\}\\
 p_3 & \textrm{otherwise}
\end{cases}\\
&=& p_3 \exp\{K_{p3}\prod_{i \neq j}\delta(m_{ij})\},\label{Pm}
\end{eqnarray}
where $\exp(K_{p2}) = \{1-(q-1)p_2\}/p_2$ and $\exp(K_{p3}) = \{1-(q^3-1)p_3\}/p_3$.
The brackets $\{\cdot\}$ express a bond set, like (12), (23), and (31) in Fig. \ref{tri}, on the elementary up-pointing triangles.

The Hamiltonian of our Potts spin glass model has the property of gauge invariance, similarly to the $\pm J$ Ising model.
The gauge transformation is defined as follows:
\begin{eqnarray}\nonumber
\phi_{ij} &\to& \phi_{ij} - (s_i - s_j) \\ \nonumber
l_{ij} &\to& l_{ij} + (s_i - s_j) \\ \nonumber
m_{ij} &\to& m_{ij} + (s_i - s_j),
\end{eqnarray}
where $s_i$ takes an integer value from $0$ to $q-1$.
The Hamiltonian is invariant under this gauge transformation.
Considering the form of the distribution functions and the property of this gauge invariance, we can calculate the exact internal energy by the well-known procedure\cite{HN81,HNbook} in a special subspace, defined by $K_2 = K_{p2} and K_3 = K_{p3}$. This subspace is usually called the Nishimori line.
Because two independent variables exist in our model, the special subspace should be called the Nishimori ``surface'' in a four-dimensional space $(K_2,K_{p2},K_3,K_{p3})$.
It is expected that there are intersections between the Nishimori surface and the phase boundary, that is, a multicritical ``line'', which consists of the multicritical points.

It is also possible to define another type of Potts spin glass with two- and three-body interactions by using the same random variables for the three-body interactions as those for the two-body interactions $\{l_{ij}\}$ instead of the random variables $\{m_{ij}\}$ in eq. (\ref{Ham}). The Hamiltonian is then written as
\begin{equation}
H = \sum_{\bigtriangleup}H_{\bigtriangleup}[\phi_{12},\phi_{23},\phi_{31}] = \sum_{\bigtriangleup}\left\{
-J_2 \sum_{i\neq j}\delta_(\phi_{ij}+l_{ij}) - J_3\prod_{i\neq j}\delta(\phi_{ij}+l_{ij})
\right\}\label{Ham2}.
\end{equation}
In this case, the following distribution function is required if we construct the Nishimori surface:
\begin{equation}
P(\{l_{ij}\}) = C \exp \left\{K_{p2}\sum_{i \neq j}\delta(l_{ij}) + K_{p3}\prod_{i \neq j}\delta(l_{ij}) \right\},\label{Pm2}
\end{equation}
where $C$ is the normalization factor.
If we define the gauge transformation as $\phi_{ij} \to \phi_{ij} - (s_i - s_j)$ and $l_{ij} \to l_{ij} + (s_i - s_j)$, the Hamiltonian is invariant under this gauge transformation.
The Nishimori surface $K_2 = K_{p2}$ and $K_3 = K_{p3}$ is also obtained.
We call the former Potts spin glass model type I, and the latter one type II.
The two slightly different spin glass models are equal to each other for the cases only with three-body interactions and only with two-body interactions.
The following analysis applies to both the type I and type II models, but only the type I case is shown.
\section{Duality and conjecture}
As used in the $\pm J$ Ising model on the triangular lattice\cite{NO}, it is straightforward to develop the direct triangular duality for the replicated system.
It is possible to apply the direct triangular duality to the replicated system as the following replicated duality relation:
\begin{eqnarray}
  \tilde{A_n^*}[\{k_{12}^{\alpha}\},\{k_{23}^{\alpha}\},\{k_{31}^{\alpha}\}]
 &=&\frac{1}{q^n}
 \sum_{\{\phi^{\alpha}_{ij}\}} \left[ \prod_{\alpha=1}^n
 \delta (\phi_{12}^{\alpha}+\phi_{23}^{\alpha}+\phi_{31}^{\alpha})\right]
  \tilde{A_n}\left[\{\phi_{12}^{\alpha}\},\{\phi_{23}^{\alpha}\},\{\phi_{31}^{\alpha}\}\right]
   \nonumber\\
 &&\times\exp \left\{i \frac{2\pi}{q}
 \sum_{\alpha=1}^n (k_1^{\alpha}\phi_{23}^{\alpha}+k_{2}^{\alpha}\phi_{31}^{\alpha}
  +k_3^{\alpha}\phi_{12}^{\alpha}) \right\},
  \label{Astar}
\end{eqnarray}
where $\alpha$ is the replica index running from 1 to $n$, and $\tilde{A}_n\left[\{\phi_{12}^{\alpha}\},\{\phi_{23}^{\alpha}\},\{\phi_{31}^{\alpha}\}\right]$ is the configurational-averaged and replicated face Boltzmann factor of the type-I Potts spin glass, which is given by
\begin{eqnarray}\nonumber
&&\tilde{A_n}\left[\{\phi_{12}^{\alpha}\},\{\phi_{23}^{\alpha}\},\{\phi_{31}^{\alpha}\}\right]\\
&&= p^3_2 p_3 \sum_{\{l_{ij},m_{ij}\}} A^{K_{p2},K_{p3}}_{\bigtriangleup}[\{l_{ij}\};\{m_{ij}\}] \prod^{n}_{\alpha=1}
A^{K_{2},K_{3}}_{\bigtriangleup}[\{ \phi^{\alpha}_{ij}+l_{ij}\};\{\phi^{\alpha}_{ij} +m_{ij}\}].\label{An}
\end{eqnarray}

Following the existing successful cases\cite{MNN,TSN,NO}, we need two conditions to derive the multicritical points:
\begin{enumerate}
	\item On the Nishimori line or, in our model, surface, the $n=1$ and $2$ replicated systems are self-dual.

We concentrate on the system on the Nishimori line, where it is believed that the multicritical point lies\cite{HN81,HNbook}.

	\item Transition points for replicated systems are determined only by a single equation:
\begin{equation}
\tilde{A_n^*}[\{0\},\{0\},\{0\}]=\tilde{A_n}[\{0\},\{0\},\{0\}].\label{conjecture}
\end{equation}

This kind of equation is satisfied in pure systems, such as the Ising model and the $q$-state Potts model on the square, triangular, and hexagonal lattices.
Indeed it is possible to obtain the fixed-point condition $y=q$ for the case of the pure Potts model with two- and three-body interactions by substituting eqs. (\ref{23P1}) and (\ref{23P2}) into eq. (\ref{conjecture}).
It is expected that eq. (\ref{conjecture}) represents the location of the singularity of the replicated system. The validity of this hypothesis is still unproven, however.
\end{enumerate}

We now examine the self-duality for the $n=1$ and $2$ replicated systems of the type-I and type-II Potts spin glass models, as discussed below.
We find that the $q$-state Potts spin glass model with only three-body interactions ($v_2=0$, $p_2 =1/q$) satisfies the self-duality for general values of $q$ when the replica numbers are $n=1$ and $2$.
On the other hand, we find that only the two-state Potts spin glass models with two- and three-body interactions, for both of the type-I and type-II cases, satisfy the self-duality when $n=1$ and $2$.
Neither the type-I nor type-II Potts spin glass model with $q\ge 3$ states satisfies the self-duality when $n=2$.

Therefore, using the conjecture, it would be possible to obtain the multicritical points of the $q$-state Potts spin glass model only with three-body interactions and both the type-I and type-II two-state Potts spin glass models with two- and three-body interactions.
On the other hand, the conjecture may give incorrect predictions for the Potts spin glass model with two- and three-body interactions with $q\ge 3$ states.
The duality is not always a good tool, even for the pure model, to predict the transition point for most systems with many states, such as the $q$-state clock model ($q\ge5$), which is not self-dual; on the other hand, the $q$-state clock model is self-dual when $q \le 4$.
This property of the duality may be the reason why the above unfavorable situation for the replicated models appears when the number of states and replicas is higher than a threshold.

The rest of this section is devoted to a detailed explanation of the above-mentioned facts about the self-duality for the replicated models.
We consider the self-duality of the type-I Potts spin glass model with two- and three-body interactions.

In the case of the $n=1$ replicated Potts spin glass model, the face Boltzmann factor is rewritten, from eq. (\ref{An}), as 
\begin{eqnarray}\nonumber
&& \tilde{A_1}\left[\phi_{12},\phi_{23},\phi_{31}\right]\\ \nonumber
&& = p_2^3p_3(q^3 + 2v_3)(q+2v_2)^3 \left[1 + \frac{v_2^2}{q+2v_2} \sum_{i \neq j}\delta(\phi_{ij})\right. \\
& & \left.+ \left\{ \frac{3v_2^4}{(q+2v_2)^2} + \frac{v_2^6}{(q+2v_2)^3} + \frac{v^2_3}{q^3+2v_3} \left(1+ \frac{v_2^2}{q+2v_2}\right)^3 \right\}\prod_{i \neq j}\delta(\phi_{ij})\right].
\end{eqnarray}
If we look at eq. (\ref{23P1}), we find that this is equivalent to the face Boltzmann factor of the Potts model with two- and three-body interactions with the effective couplings $\tilde{v_2} = v_2^2/(q+2v_2)$, and $\tilde{v_3} = v^2_3/(q^3 + 2v_3)$.
It is therefore easy to obtain the following fixed-point condition, similarly to the case when we obtained the fixed-point condition (\ref{dr}):
\begin{equation}
3\tilde{v}_2^2 + \tilde{v}^3_2 + \tilde{v}_3(1+\tilde{v}_2)^3 = q. \label{fc1}
\end{equation}

In the case of the $n=2$ replicated Potts spin glass model, the face Boltzmann factor is rewritten, from eq. (\ref{An}), as:  
\begin{eqnarray}\nonumber
&&\tilde{A_2}\left[\{\phi^{\alpha}_{12}\},\{\phi^{\alpha}_{23}\},\{\phi^{\alpha}_{31}\}\right]\\ \nonumber
&& = p_3 p_2^3
\prod_{i \neq j} \left\{q+3v_2 + v_2^2\sum^{2}_{\alpha=1}\delta(\phi^{\alpha}_{ij})+ v_2^2\delta(\phi^1_{ij}-\phi^2_{ij}) + v_2^3\prod^2_{\alpha=1}\delta(\phi^{\alpha}_{ij})  \right\}
\\ 
&& \times \left\{q^3+3v_3 + v_3^2\sum^{2}_{\alpha=1}\prod_{i \neq j}\delta(\phi^{\alpha}_{ij})+ v_3^2\prod_{i \neq j}\delta(\phi^1_{ij}-\phi^2_{ij}) + v_3^3\prod^2_{\alpha=1}\prod_{i \neq j}\delta(\phi^{\alpha}_{ij})  \right\}.\label{23Pottsn=2}
\end{eqnarray}
This face Boltzmann factor does not generally show the self-dual property.
The reason is that, in eq. (\ref{23Pottsn=2}), the first product over $i \neq j$ produces the products of Kronecker's delta symbols, such as $\delta(\phi^{1}_{12})\delta(\phi_{23}^{2})\delta(\phi^1_{31}-\phi^2_{31})$, and its cyclic permutations of the indices.
The dual terms of these products are given by the application of eq. (\ref{Astar}).
For example, $\delta(\phi^{1}_{12})\delta(\phi_{23}^{2})\delta(\phi^1_{31}-\phi^2_{31})$ is transformed as follows:
\begin{equation}
\delta(\phi^{1}_{12})\delta(\phi_{23}^{2})\delta(\phi^1_{31}-\phi^2_{31})
\to \frac{1}{q}\delta(k^1_{12} - k^2_{23}).\label{difficult}
\end{equation}
However, this type of interaction does not exist in eq. (\ref{23Pottsn=2}). 
Therefore, the $n=2$ replicated Potts spin glass model with two- and three-body interactions is not generally self-dual.
We thus conclude that the condition (i) is not always satisfied for the Potts spin glass model with two- and three-body interactions.
We find also that the $q$-state Potts spin glass model only with two-body interactions does not satisfy the condition (i) either, because this problem is produced by the first product over $i \neq j$ in eq. (\ref{23Pottsn=2}), which is obtained from terms representing two-body interactions in eq. (\ref{An}).
On the other hand, when we consider the case $v_2 = 0$ ($p_2 = 1/q)$, only with three-body interactions, such troublesome terms as $\delta(\phi^{1}_{12})\delta(\phi_{23}^{2})\delta(\phi^1_{31}-\phi^2_{31})$  vanish.
More precisely, the face Boltzmann factor (\ref{23Pottsn=2}) is reduced to
\begin{eqnarray}\nonumber
&&\tilde{A_2}\left[\{\phi^{\alpha}_{12}\},\{\phi^{\alpha}_{23}\},\{\phi^{\alpha}_{31}\}\right]\\ \nonumber
&&= p_3 (q^3+3v_3)\left\{1 + \frac{v_3^2}{q^3 + 3v_3}\sum^{2}_{\alpha=1}\prod_{i \neq j}\delta(\phi^{\alpha}_{ij}) \right.\\
&& \left.+ \frac{v_3^2}{q^3 + 3v_3}\prod_{i \neq j}\delta(\phi^1_{ij}-\phi^2_{ij})  + \frac{v_3^3}{q^3 + 3v_3}\prod^2_{\alpha=1}\prod_{i \neq j}\delta(\phi^{\alpha}_{ij})  \right\}.\label{3Pottsn=2}
\end{eqnarray}
Though this face Boltzmann factor is still a little complicated, the terms involving Kronecker's delta symbols, such as $\prod_{i \neq j} \delta(\phi^{\alpha}_{ij})$, and the constant $1$ are changed in the following manner by the application of the direct triangular duality (\ref{Astar}):
\begin{equation}\nonumber
\begin{array}{rclrcl}
1 &\to& q^2 \displaystyle \prod^2_{\alpha=1} \delta(k^{\alpha}_{ij}) & \delta(\phi^{\alpha}_{ij}) &\to& \delta(k^{\beta}_{ij}) \quad (\beta \neq \alpha)\\
\delta(\phi^{1}_{ij}-\phi^{2}_{ij}) &\to& \delta(k^{1}_{ij}-k^{2}_{ij}) & \displaystyle \prod^2_{\alpha=1} \delta(\phi^{\alpha}_{ij}) &\to& \displaystyle \frac{1}{q^2}.
\end{array}
\end{equation}
From these transformations, the explicit form of the dual face Boltzmann factor is
\begin{eqnarray}\nonumber
&&\tilde{A_2^*}\left[\{k^{\alpha}_{12}\},\{k^{\alpha}_{23}\},\{k^{\alpha}_{31}\}\right]\\
&&=\frac{p_3 v_3^3}{q^2}\left\{1 + \frac{q^2}{v_3}\sum^{2}_{\alpha=1}\prod_{i \neq j}\delta(k^{\alpha}_{ij}) + \frac{q^4(q^3 + 3 v_3)}{v^3_3}\prod_{i \neq j}\delta(k^1_{ij}-k^2_{ij})  + \frac{q}{v_3}\prod^2_{\alpha=1}\prod_{i \neq j}\delta(k^{\alpha}_{ij})\right\}.\nonumber \\
\end{eqnarray}
Therefore, the $n=2$-replicated $q$-state Potts spin glass only with three-body interactions is self-dual.
Thus, it is confirmed that the condition (i) is satisfied for the Potts spin glass model only with three-body interactions.
The fixed-point condition is expressed as a single equation:
\begin{equation}
\frac{v_3^2}{q^3+3v_3} = q^2.\label{fc2}
\end{equation}

If we restrict ourselves to the case $q=2$, which is equivalent to the Ising system, we find a possibility that the system is self-dual.
Such terms as $\delta(k^1_{12}-k^2_{23})$ are rewritten as follows, only in the $q=2$ case:
\begin{equation}
\delta(k^1_{12}-k^2_{23}) = 2\delta(k^1_{12})\delta(k^2_{23})-\delta(k^1_{12})-\delta(k^2_{23}) + 1.
\end{equation}
The terms on the right-hand side can be obtained from eq. (\ref{23Pottsn=2}).
Terms consisting of the products of the Kronecker's delta symbols not appearing in eq. (\ref{23Pottsn=2}) are obtained by the direct triangular duality (\ref{Astar}).
The $n=2$-replicated $2$-state Potts spin glass is therefore self-dual.
In Appendix A, this self-duality is shown more explicitly than here, and the fixed-point condition for the two-state Potts spin glass model is derived.
The condition (i) is also satisfied by the two-state Potts spin glass model.

We conclude that it is possible to predict two series of the locations of the multicritical points: the multicritical points of the general $q$-state Potts spin glass model only with three-body interactions and the multicritical lines of both the type-I and type-II two-state Potts spin glass models with two- and three-body interactions.

\section{Multicritical points and lines}
Our next tasks are to confirm that the condition (ii) is satisfied and to predict the multicritical points and lines.
It is possible to write the following explicit form of eq. (\ref{conjecture}) for the $q$-state Potts spin glass model, as shown in Appendix B: 
\begin{eqnarray}\nonumber
& & p^3_2 p_3 
\left\{
q-1 + (1 + v_2)^{n+1}
\right\}^3
\left\{
q^3-1 + (1 + v_3)^{n+1}
\right\}^3
\\
& & = \frac{p^3_2 p_3}{q^n}
\left\{
q^2(q-1)^2 F^{n+1} + q^2(q-1) G^{n+1} 
+ \sum^3_{i=0}\left(a_i F^{n+1}_i + b_i G^{n+1}_i\right)
\right\}
. \label{23MCPI}
\end{eqnarray}
where 
\begin{equation}\nonumber
\begin{array}{rclrcl}
F &=& \displaystyle \frac{1}{q}\left\{(q+v_2)^3 - v_2^3 \right\} & F_i &=& F + v_3(1+v_2)^i \\
& & & & & \\
G &=& \displaystyle \frac{1}{q}\left\{(q+v_2)^3 + (q - 1) v_2^3 \right\} & G_i &=& G + v_3(1+v_2)^i ,
\end{array}
\end{equation}
and $a_i$ and $b_i$ are the coefficients of $F_i$ and $G_i$, which are evaluated as follows:
\begin{equation}\nonumber
\begin{array}{rclrcl}
a_0 &=& (q-1)^3 +(q-1)(q-2)  & b_0&=& (q-1)(q-2)\\
a_1 &=& 3(q-1)(q-2) & b_1&=& 3(q-1) \\
a_2 &=& 3(q-1) & b_2&=& 0 \\ 
a_3 &=& 0 & b_3&=& 1.
\end{array}
\end{equation}
Detailed calculations are shown in Appendix B.
Substituting $n=1$ and $n=2$ with $q=2$ into eq. (\ref{23MCPI}), we obtain the same equations as the fixed-point conditions for the $n=1$- and $n=2$-replicated Potts spin glasses with two states.
When $v_2=0$ $(p_2 = 1/q)$, eq. (\ref{23MCPI}) also gives the fixed-point conditions (\ref{fc1}) and (\ref{fc2}).
It is confirmed that the condition (ii) is also satisfied for the $q$-state Potts spin glass model only with three-body interactions and the two-state Potts spin glass model.
In accordance with the replica method, we consider the $n \to 0$ limit of eq. (\ref{23MCPI}), which is expected to determine the singularity of the Potts spin glass model.
We obtain the following equation, which gives predictions of the multicritical points for various combinations $p_2$ and $p_3$, by taking the leading term of $n$ in eq. (\ref{23MCPI}):
\begin{eqnarray}\nonumber
& &
q^2(q-1)^2 F\ln F + q^2(q-1) G\ln G 
+ \sum^3_{i=0}\left(a_i F_i \ln F_i + b_i G_i \ln G_i\right) 
\\ \nonumber
& & \quad -(1+v_3)(q+v_2)^3\ln(1 + v_3)
- 3(q+v_2)^2(1+v_2)(q^3 + v_3)\ln(1 + v_2)
\\ && \qquad= (q^3 + v_3)(q+v_2)^3\ln q
.  \label{23MCPI2}
\end{eqnarray}
The solutions to this equation will describe a multicritical line on the $(p_2, p_3)$ or $(K_{p2}, K_{p3})$ plane.
The above result is for the type-I Potts spin glass.
We also derive the following equation of the multicritical line for the type-II Potts spin glass:
\begin{eqnarray}\nonumber
& &
(q-1)F\ln F + G_3 \ln G_3 - 3v_3(1+v_2)^3\ln(1 + v_2)
- 3(q+v_2)^2(1+v_2)\ln(1 + v_2)
\\
& &
-(1+v_3)(1+v_2)^3\ln(1 + v_3) = \left\{ (q-1)F + G_3\right\} \ln q
. \label{23MCPII}
\end{eqnarray}
When we consider the case of $v_2=0$ ($p_2 = 1/q$), eq. (\ref{23MCPI2}) gives the following equation, which determines the location of the multicritical points of the $q$-state Potts spin glass model only with the three-body interactions:
\begin{eqnarray}\nonumber
& & \left\{1 - q^2(q-1)p_3\right\} \ln \left\{1 - q^2(q-1)p_3\right\} 
- \left\{1 - (q^3-1)p_3\right\} \ln \left\{1 - (q^3-1)p_3\right\} \\
& & -(q^3-1)p_3 \ln p_3 = \left\{ 1-2q^2(q-1)p_3 \right\} \ln q.\label{3MCP2}
\end{eqnarray}
The solutions are, for example, $p_{3c} = 0.0396336$ ($q=2$), $p_{3c}=0.0139431$ ($q=3$), $p_{3c} = 0.00637851$ ($q=4$), and $p_{3c}= 0.00342033$ ($q=5$).

\begin{figure}[tb]
\begin{center}
\includegraphics[width=70mm]{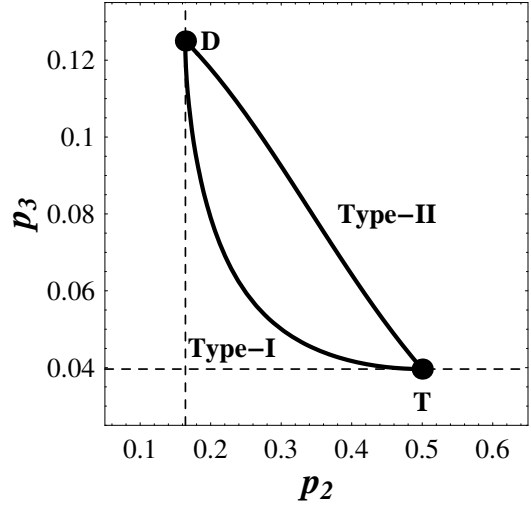}
\end{center}
\caption{{\small The two solid lines connecting $(p_2, p_3) = (1/2, 0.0396336)$ and $(p_2,p_3) = (0.164194, 1/2^3)$ are the multicritical lines. The two points T and D are the multicritical points of the Potts spin glass model only with three-body interactions ($K_2 = 0$, $p_2 = 1/q$) and only with two-body interactions ($K_3=0$, $p_3=1/q^3$), respectively. }
}
\label{MCLs}
\end{figure}

Both lines described by eqs. (\ref{23MCPI2}) and (\ref{23MCPII}) are shown in Fig. \ref{MCLs}.
This is a reasonable result because both lines smoothly connect the multicritical point of the $2$-state Potts spin glass model only with three-body interactions $p_{3c}=0.03969336$ with that of the $2$-state Potts spin glass model only with two-body interactions $p_{2c} = 0.164194$, which is equivalent to the Ising model on the triangular lattice\cite{NO}. Here, $p_2$ expresses the density of the antiferromagnetic interactions.
\section{Discussion}
We introduced two Potts spin glass models, type-I and type-II, on the triangular lattice.
Similarly to the $\pm J$ Ising model, it is possible to find a special subspace of the phase diagram where we can calculate the exact internal energy, known as the Nishimori line.
The special subspace should be a surface, because the Potts spin glass models have two coupling parameters.
It is expected that, considering the properties of the Nishimori line, the multicritical points lie on a surface.
It is also expected that the multicritical points describe a line, because the multicritical points are located at the intersections between the phase boundary and this surface.
Applying the direct triangular duality in conjunction with the replica method to the two types of Potts spin glass models, we argued for the self-duality of the replicated systems of these models.
Taking account of the existing successful examples of the conjecture for many spin glass models, we expect that the condition to obtain the exact location of the multicritical points is the self-duality of the replicated models with $n = 1$ and $n = 2$ replicas.
We showed the self-duality in only the cases of the $q$-state Potts spin glass model with only three-body interactions and two types of $2$-state Potts spin glass models with two- and three-body interactions.
Thus, our results possibly give the exact location of the multicritical points of the general Potts spin glass model with only three-body interactions with $q$ states and the multicritical lines of two slightly different Potts spin glass models for the $2$-state case.
However, the hypothesis that a single equation will determine the singularity of the spin glass models is not completely verified.
Therefore, we have to compare our results with the results obtained by other techniques.

On the other hand, neither of the $n=2$ replicated Potts spin glass models with two- and three-body interactions with $q\ge 3$ states satisfy the self-duality, which may result in incorrect predictions for this case.
However, it is not certain which conditions are truly necessary.
It would also be worth investigating the location of the multicritical lines of the Potts spin glass models with $q\ge 3$ states to verify the necessity of the self-duality for the $n=2$-replicated Potts spin glass model.

For the above reasons, it is expected that our multicritical lines with $2$ states will be helpful to further investigate the validity of the conjecture.
Moreover, an investigation of the phase transition of the Potts spin glass model with $q\ge 3$ states by other approaches should elucidate the precise conditions necessary to obtain the multicritical points.
\section{Acknowledgements}
The author would like to thank Prof. H. Nishimori of the Tokyo Institute of Technology for useful discussions and suggestions on the manuscript. This work was partially supported by the 21st Century COE Program `Nanometer-Scale Quantum Physics' at the Tokyo Institute of Technology, and by Grant-in-Aid for Scientific Research on the Priority Area ``Deepening and Expansion of Statistical Mechanical Informatics'' from the Ministry of Education, Culture,
Sports, Science, and Technology of Japan and by CREST, JST.

\appendix
\section{Duality of the $n=2$-replicated Potts spin glass model with $2$ states}
It is a little complicated to write the expansion of eq. (\ref{23Pottsn=2}). Therefore, we directly calculate the face Boltzmann factors considering symmetry in eq. (\ref{23Pottsn=2}). Because of the constraint $\phi^{\alpha}_{12}+\phi^{\alpha}_{23}+\phi^{\alpha}_{31} \equiv 0$ with mod $2$, the following four types of face Boltzmann factors from eq. (\ref{23Pottsn=2}) characterize the replicated Potts spin glass model:
\begin{eqnarray}\nonumber
A_0 &\equiv& \tilde{A_2}[0,0,0;0,0,0]\\
&=& p_2^3p_3 
\left(
q + 3v_2 +3v_2^2+v_2^3
\right)^3
\left(
q^3 + 3v_3 +3v_3^2+v_3^3
\right)\\\nonumber
A_1 &\equiv& \tilde{A_2}[0,0,0;0,1,1] = \tilde{A_2}[0,0,0;1,0,1] =\tilde{A_2}[0,0,0;1,1,0]\\\nonumber
&=&\tilde{A_2}[0,1,1;0,0,0]=\tilde{A_2}[1,0,1;0,0,0]=\tilde{A_2}[1,1,0;0,0,0]\\
&=& p_2^3p_3 
\left(
q + 3v_2 +3v_2^2+v_2^3
\right)
\left(
q + 3v_2 +v_2^2
\right)^2
\left(
q^3 + 3v_3 +v_3^2
\right)\\\nonumber
A_2 &\equiv& \tilde{A_2}[0,1,1;0,1,1] = \tilde{A_2}[1,0,1;1,0,1] =\tilde{A_2}[1,1,0;1,1,0]\\
&=& p_2^3p_3 
\left(
q + 3v_2 +3v_2^2+v_2^3
\right)
\left(
q + 3v_2 +v_2^2
\right)^2
\left(
q^3 + 3v_3
\right)\\\nonumber
A_3 &\equiv& \tilde{A_2}[0,1,1;1,0,1] = \tilde{A_2}[0,1,1;1,1,0] = \tilde{A_2}[1,0,1;0,1,1] = \tilde{A_2}[1,0,1;1,1,0]\\\nonumber
&=&\tilde{A_2}[1,1,0;0,1,1] = \tilde{A_2}[1,1,0;1,0,1]\\
&=& p_2^3p_3 
\left(
q + 3v_2 +v_2^2
\right)^3
\left(
q^3 + 3v_3
\right).
\end{eqnarray}
Using these face Boltzmann factors, it is possible to write its dual face Boltzmann factor using eq. (\ref{Astar}), as follows:
\begin{eqnarray}\nonumber
&& \tilde{A^*_2}[\{k^{\alpha}_{12}\},\{k^{\alpha}_{23}\},\{k^{\alpha}_{31}\}]\\\nonumber
&=& \frac{1}{4}\left[A_0 + A_1\sum^2_{\alpha=1}\sum_{i \neq j}\exp \left(i \pi k^{\alpha}_{ij}\right) \right.\\ 
& & \left.+ A_2 \sum_{i\neq j}\exp \left\{i \pi (k^1_{ij}+k^2_{ij})\right\} + A_3 \sum_{ijk}\exp \left\{i \pi (k^1_{jk}+k^2_{ki})\right\}
\right].
\end{eqnarray}
This face Boltzmann factor has the same symmetry as the original one $\tilde{A_2}[\{\phi^{\alpha}_{12}\},\{\phi^{\alpha}_{23}\},\{\phi^{\alpha}_{31}\}]$; for example, $\tilde{A^*_2}[0,0,0;0,1,1]= \tilde{A^*_2}[0,0,0;1,0,1] =\tilde{A^*_2}[0,0,0;1,1,0] = \tilde{A^*_2}[0,1,1;0,0,0]=\tilde{A^*_2}[1,0,1;0,0,0]=\tilde{A^*_2}[1,1,0;0,0,0]$.
Therefore, the following duality relations are obtained:
\begin{eqnarray}
A_0^* &=& \frac{1}{4}\left(A_0 + 6A_1 + 3A_2 + 6A_3\right)\\
A_1^* &=& \frac{1}{4}\left(A_0 + 2A_1 - A_2 - 2A_3\right)\\
A_2^* &=& \frac{1}{4}\left(A_0  -2A_1 + 3A_2 - 2A_3\right)\\
A_3^* &=& \frac{1}{4}\left(A_0  -2A_1 - 1A_2 + 2A_3\right).\
\end{eqnarray}
Setting $\omega_i \equiv A_i/A_0$, the above equations are rewritten as
\begin{equation}
\frac{\omega^*_i +\displaystyle\frac{1}{3}}{\omega_i +\displaystyle\frac{1}{3}}= \frac{4}{1 + 6\omega_1 + 3\omega_2 + 6\omega_3}.
\end{equation}
It is therefore possible to obtain the following fixed condition, where $\omega_i = \omega^*_i$ for any $i$: 
\begin{equation}
1 + 6\omega_1 + 3\omega_2 + 6\omega_3 = 4.
\end{equation}
This condition can again be derived by $A_0 = A_0^*$, that is, eq. (\ref{conjecture}).
It is also straightforward to show that the general cases of $q$ states are not self-dual, similarly to the above discussion, or by direct manipulation of eq. (\ref{Astar}).

\section{Evaluation of eq.(\ref{23MCPI})}
We will show the evaluation of eq. (\ref{23MCPI}) in this section.
The quantity of eq. (\ref{23MCPI}) is obtained by substituting $\{\phi^{\alpha}_{ij}\} = \{0\}$ into eq. (\ref{An}) and by substituting $\{k^{\alpha}_{ij}\} = \{0\}$ into eq. (\ref{Astar}) under the Nishimori surface conditions $K_2 = K_{p2}$, and $K_3 = K_{p3}$.

First, the left-hand side of eq. (\ref{23MCPI}) is obtained from eq. (\ref{An}), as follows:
\begin{eqnarray}\nonumber
&&\tilde{A_n}\left[\{0\},\{0\},\{0\}\right]\\ \nonumber
&&= p^3_2 p_3 \sum_{\{l_{ij}\}} \sum_{\{m_{ij}\}}A^{K_{2},K_{3}}_{\bigtriangleup}[\{l_{ij}\};\{m_{ij}\}] \times \prod^{n}_{\alpha=1}
A^{K_{2},K_{3}}_{\bigtriangleup}[\{l_{ij}\};\{m_{ij}\}]\\ \nonumber
&&= p^3_2 p_3 \sum_{\{l_{ij}\}} \sum_{\{m_{ij}\}} \prod_{i \neq j}
\exp
\left\{
(n+1)K_2 \delta(l_{ij}) 
\right\}
\exp
\left\{
(n+1)K_3 \prod_{i \neq j} \delta(m_{ij})
\right\}
\\ \nonumber
&&= p^3_2 p_3 \prod_{i \neq j}
\left[
\sum_{\{l_{ij}\}}
\left\{ 1 + (e^{(n+1)K_2}-1) \delta(l_{ij})
\right\}
\right]
\left[
\sum_{\{m_{ij}\}}
\left\{ 1 + (e^{(n+1)K_3}-1) \prod_{ij}\delta(m_{ij})
\right\}
\right]
\\
&&= p^3_2 p_3
\left\{ q -1 + (1 + v_2)^{n+1}
\right\}^3
\left\{ q^3 -1  + (1+v_3)^{n+1}
\right\}.
\end{eqnarray}

Next, we derive the expression on the right-hand side of eq. (\ref{23MCPI}).
We substitute $\{k^{\alpha}_{ij}\} = \{0\}$, that is $\{k^{\alpha}_i\} = \{k^{\alpha}_i\}$, into eq. (\ref{Astar}) under the conditions of the Nishimori surface. Then, the exponential term in eq. (\ref{Astar}) vanishes because of the constraints by Kronecker's delta $\delta (\phi_{12}^{\alpha}+\phi_{23}^{\alpha}+\phi_{31}^{\alpha})$ for each replica.
It is therefore necessary to evaluate only the following quantity:
\begin{eqnarray}\nonumber
 && \tilde{A_n^*}[\{0\},\{0\},\{0\}] \\ \nonumber
 && = \frac{p^3_2 p_3}{q^n} \sum_{\{l_{ij}\}} \sum_{\{m_{ij}\}} A^{K_{2},K_{3}}_{\bigtriangleup}[\{l_{ij}\};\{m_{ij}\}] \\ 
 && \quad \times
   \prod_{\alpha=1}^n  \left[ \sum_{\{\phi_{ij}\}}
 \delta (\phi_{12}^{\alpha}+\phi_{23}^{\alpha}+\phi_{31}^{\alpha})
A^{K_{2},K_{3}}_{\bigtriangleup}[\{ \phi^{\alpha}_{ij}+l_{ij}\};\{\phi^{\alpha}_{ij} +m_{ij}\}]\right]. \label{1Astar}
\end{eqnarray}
We sum over $\phi_{ij}$ for later convenience, as follows:
\begin{eqnarray}\nonumber
& &  \prod_{\alpha=1}^n \left[ \sum_{\{\phi_{ij}\}}
 \delta (\phi_{12}^{\alpha}+\phi_{23}^{\alpha}+\phi_{31}^{\alpha})
A^{K_{2},K_{3}}_{\bigtriangleup}[\{ \phi^{\alpha}_{ij}+l_{ij}\};\{\phi^{\alpha}_{ij} +m_{ij}\}] \right] \\ \nonumber
& &  = \prod_{\alpha=1}^n \left[ \sum_{\{\phi_{ij}\}}
 \delta (\phi_{12}^{\alpha}+\phi_{23}^{\alpha}+\phi_{31}^{\alpha})
\prod_{i \neq j}\{1 + v_2 \delta(\phi_{ij}+l_{ij})\} \{1 + v_3 \prod_{i\neq j}\delta(\phi_{ij} + m_{ij})\} \right] \\ 
& &  = F + C(\{l_{ij}\};\{m_{ij}\}), \label{resum}
\end{eqnarray}
where the two quantities $F$ and $C(\{l_{ij}\}:\{m_{ij}\})$ are defined as follows:
\begin{eqnarray}
F &\equiv& \frac{1}{q}\left\{(q+v_2)^3 -v_2^3\right\}\\ \nonumber
C(\{l_{ij}\};\{m_{ij}\}) &\equiv& v_3 \delta(m_{12}+m_{23}+m_{31})
\prod_{i \neq j} 
\{1 + v_2 \delta(l_{ij}-m_{ij})\} 
+ v^3_2 \delta(l_{12}+l_{23}+l_{31}). \\
\end{eqnarray}
Substituting eq. (\ref{resum}) into eq. (\ref{1Astar}) yields
\begin{eqnarray}\nonumber
& &\tilde{A_n^*}[\{0\},\{0\},\{0\}] \\ \nonumber
& & = p_2^3 p_3 v_3\sum_{\{l_{ij}\}} \prod_{i\neq j} 
\left\{ 
1 + v_2 \delta(l_{ij})
\right\}
\left\{
F + C(\{l_{ij}\};\{0\})
\right\}^n \\ 
& & \quad + p_2^3 p_3 \sum_{\{l_{ij}\}}\sum_{\{m_{ij}\}}\prod_{i\neq j} 
\left\{ 
1 + v_2 \delta(l_{ij})
\right\}
\left\{
F + C(\{l_{ij}\};\{m_{ij}\})
\right\}^n.
\end{eqnarray}
We divide both sets $\{l_{ij}\}$ and $\{m_{ij}\}$ into two subsets $P$ and $\bar{P}$ to sum over $\{l_{ij}\}$ and $\{m_{ij}\}$.
Here, $P$ is a subset which satisfies the constraints $x_{12}+x_{23}+x_{31}\neq 0 $(mod $q$), $x_{ij} \in P$.
The reason why we consider such division of sets $\{l_{ij}\}$ and $\{m_{ij}\}$ is that the quantity $C(\{l_{ij}\};\{m_{ij}\})$ depends on the values of $l_{12}+l_{23}+l_{31}$ and $m_{12}+m_{23}+m_{31}$.
Dividing sets $\{l_{ij}\}$ and $\{m_{ij}\}$ and expanding the terms with power $n$ yields
\begin{eqnarray}\nonumber
& &\tilde{A_n^*}[\{0\},\{0\},\{0\}] \\ \nonumber
&& = 
\frac{p_2^3p_3 v_3}{q^n}
\left[
\sum_{\{l_{ij}\} \in P}
 \sum_{r=0}^{n}\left(
\begin{array}{c}
n \\
r 
\end{array}
\right)
F^{n-r} v_3^r \prod_{i \neq j}\exp \left\{ (r+1)K_2 \delta(l_{ij})\right\}
\right.
\\ \nonumber
&& \left.
\quad + \sum_{\{l_{ij}\}\in \bar{P}}
\sum_{r=0}^{n}\left(
\begin{array}{c}
n \\
r 
\end{array}
\right)
\left(F+v_2^3\right)^{n-r} v_3^r \prod_{i \neq j}\exp \left\{ (r+1)K_2 \delta(l_{ij})\right\}
\right]
\\ \nonumber
&& \quad + 
\frac{p_2^3p_3}{q^n}\left[
\sum_{\{l_{ij}\}\in P} \sum_{\{m_{ij}\}\in P}
 F^n \prod_{i \neq j}
\exp \left\{ K_2 \delta(l_{ij})\right\}
\right.
\\ \nonumber
&& \quad + \sum_{\{l_{ij}\}\in \bar{P}} \sum_{\{m_{ij}\}\in P}
 \prod_{i \neq j}
\exp \left\{ K_2 \delta(l_{ij})\right\}
\\ \nonumber
&& \quad + \sum_{\{l_{ij}\}\in P}\sum_{\{m_{ij}\}\in \bar{P}}
 \sum_{r=0}^{n}\left(
\begin{array}{c}
n \\
r 
\end{array}
\right)
F^{n-r} v_3^r \prod_{i \neq j}\exp \left\{ K_2 \delta(l_{ij}) + rK_2 \delta(l_{ij}-m_{ij})\right\}
\\ \nonumber
&& \quad + \left.\sum_{\{l_{ij}\}\in \bar{P}}\sum_{\{m_{ij}\}\in \bar{P}}
 \sum_{r=0}^{n}\left(
\begin{array}{c}
n \\
r 
\end{array}
\right)
\left(F+v_2^3\right)^{n-r} v_3^r \prod_{i \neq j}\exp \left\{ K_2 \delta(l_{ij}) + rK_2 \delta(l_{ij}-m_{ij})\right\}
\right].
\\ \label{resum2}
\end{eqnarray}
It is possible to evaluate all terms except for the last two terms in this equation as follows:
\begin{equation}
\frac{p_2^3p_3 }{q^n}
\left[v_3
\sum_{i=0}^{3}\left\{
a_i 
F_i^n
(1+v_2)^i +b_i 
G_i^n
(1+v_2)^i\right\}
+
q^2(q-1)^2 F^{n+1} + 
q^2(q-1) 
G^{n+1}
\right], \label{resum3}
\end{equation}
where 
\begin{equation}
\begin{array}{rclrcl}
a_0 &=& (q-1)^3 +(q-1)(q-2)  & b_0&=& (q-1)(q-2)\\
a_1 &=& 3(q-1)(q-2) & b_1&=& 3(q-1) \\
a_2 &=& 3(q-1) & b_2&=& 0 \\ 
a_3 &=& 0 & b_3&=& 1 .\\
\end{array}
\end{equation}
Here, we used the following convenient identities:
\begin{eqnarray}
\sum_{\{l_{ij}\}} \prod_{i \neq j} \exp \left\{ R \delta(l_{ij})\right\}
&=& (q + v_R)^3 \\
\sum_{\{l_{ij}\} \in \bar{P}} \prod_{i \neq j} \exp \left\{ R \delta(l_{ij})\right\}
&=& \frac{1}{q}\left\{(q + v_R)^3 +(q-1)v_R^3\right\} \equiv G \\
\sum_{\{l_{ij}\} \in P} \prod_{i \neq j} \exp \left\{ R \delta(l_{ij})\right\}
&=& \left(1-\frac{1}{q}\right)\left\{(q + v_R)^3 -v_R^3\right\} = (q-1)F.
\end{eqnarray}
On the other hand, the last two terms of eq. (\ref{resum2}) are evaluated as follows, considering the relationship of the two subsets $P$ and $\bar{P}$:
\begin{eqnarray}\nonumber
&& \frac{p_2^3p_3}{q^n} \left[\sum_{\{l_{ij}\}\in P}\sum_{\{m_{ij}\}\in \bar{P}}
 \sum_{r=0}^{n}\left(
\begin{array}{c}
n \\
r 
\end{array}
\right)
F^{n-r} v_3^r \prod_{i \neq j}\exp \left\{ K_2 \delta(l_{ij}) + rK_2 \delta(l_{ij}-m_{ij})\right\}\right.
\\ \nonumber
&& \left.\quad + \sum_{\{l_{ij}\}\in \bar{P}}\sum_{\{m_{ij}\}\in \bar{P}}
 \sum_{r=0}^{n}\left(
\begin{array}{c}
n \\
r 
\end{array}
\right)
\left(F+v_2^3\right)^{n-r} v_3^r \prod_{i \neq j}\exp \left\{ K_2 \delta(l_{ij}) + rK_2 \delta(l_{ij}-m_{ij})
\right\}
\right]
\\ \nonumber
&& = \frac{p_2^3p_3}{q^n} \left[
\sum_{\{l_{ij}\}}\sum_{\{m_{ij}\}\in \bar{P}}
 \sum_{r=0}^{n}\left(
\begin{array}{c}
n \\
r 
\end{array}
\right)
F^{n-r} v_3^r \prod_{i \neq j}\exp \left\{ K_2 \delta(l_{ij}) + rK_2 \delta(l_{ij}-m_{ij})\right\}\right.
\\ \nonumber
&& \quad + \left. \sum_{\{l_{ij}\}\in \bar{P}}\sum_{\{m_{ij}\}\in \bar{P}}
 \sum_{r=0}^{n}\left(
\begin{array}{c}
n \\
r 
\end{array}
\right)
\left\{ \left(F+v_2^3\right)^{n-r}-F^{n-r}\right\} v_3^r \prod_{i \neq j}\exp \left\{ K_2 \delta(l_{ij}) + rK_2 \delta(l_{ij}-m_{ij})
\right\}
\right].
\\ \label{resum4}
\end{eqnarray}
We use another expression for limitation of the summation over $\{l_{ij}\} \in \bar{P}$ and $\{m_{ij}\} \in \bar{P}$:
\begin{eqnarray}\nonumber
\sum_{\{l_{ij}\} \in \bar{P}} &=& \sum_{\{l_{ij}\}}\delta(l_{12}+l_{23}+l_{31})\\ \nonumber
&=& \frac{1}{q}\sum_{\{l_{ij}\}}\sum^{q-1}_{\tau_l=0}\exp
\left\{
i \frac{2\pi}{q}(l_{12}+l_{23}+l_{31})\right\}.
\end{eqnarray}
Eq. (\ref{resum4}) is then evaluated as follows
\begin{eqnarray}\nonumber
& & \frac{p_2^3p_3}{q^n}\left[\frac{1}{q}\sum_{\tau_l = 0}^{q-1}
\sum_{r=0}^{n}\left(
\begin{array}{c}
n \\
r 
\end{array}
\right)
F^{n-r} v_3^r\left\{ q(q+v_2 + v^{(r)}_2) \delta(\tau_l) + v_2 v_2^{(r)}\right\}^3 \right.
\\ \nonumber
& & \quad +
\frac{1}{q^2}\sum_{\tau_l = 0}^{q-1}\sum_{\tau_m = 0}^{q-1}
\sum_{r=0}^{n}\left(
\begin{array}{c}
n \\
r 
\end{array}
\right)
\left\{(F+v_2^3)^{n-r} - F^{n-r} \right\}v_3^r
\\
& & \left.\quad \quad \times
\left\{ q^2 \delta(\tau_l)\delta(\tau_m) + qv_2 \delta(\tau_l) + qv_2^{(r)}\delta(\tau_l+\tau_m)v_2 + v_2^{(r)}\right\}^3\right],
\end{eqnarray}
where $v_2^{(r)} \equiv \exp(rK_2)-1$.
It is straightforward to expand this equation and obtain the following expression:
\begin{equation}
\frac{p_2^3p_3}{q^n}\left[
\sum_{i=0}^3 a_i F
F_i^n
+
\sum_{i=0}^3 b_i G
G_i^n
\right].\label{resum5}
\end{equation}
Adding eq. (\ref{resum3}) to eq. (\ref{resum5}), we obtain the following result for $\tilde{A}_n^*[\{0\},\{0\},\{0\}]$, which is the right-hand side of eq. (\ref{23MCPI}):
\begin{equation}
\tilde{A}_n^*[\{0\},\{0\},\{0\}] = \frac{p_2^3p_3}{q^n}\left[
q^2(q-1)^2 F^{n+1} + q^2 (q-1) G^{n+1}
+
\sum_{i=0}^3 a_i F_i^{n+1}
+
\sum_{i=0}^3 b_i G_i^{n+1}
\right].\label{resum6}
\end{equation}


\end{document}